\documentclass[10pt,a4paper]{article}

\usepackage{amsmath,amsthm,amsfonts}
\usepackage{latexsym}
\usepackage{setspace}
\usepackage{natbib}
\usepackage{verbatim}
\usepackage{epstopdf}
\usepackage{epsfig}
\usepackage{lscape}
\usepackage{float}
\usepackage[T1]{fontenc}
\usepackage[utf8]{inputenc}
\usepackage{authblk}

\restylefloat{table}

\begin{document}

\title{The Excess Returns of ``Quality'' Stocks: A Behavioral Anomaly}

\author[1]{Jean-Philippe Bouchaud} 
\author[2]{Stefano Ciliberti} 
\author[3]{Augustin Landier} 
\author[2]{Guillaume Simon}  
\author[4]{David Thesmar}

\affil[1]{Capital Fund Management,\footnote{23 rue de l'Universit\'e, 75007 Paris, France} and 
CFM-Imperial Institute of Quantitative Finance \footnote{Department of Mathematics, Imperial College, 
180 Queen's Gate, London SW7 2RH} } 
\affil[2]{Capital Fund Management}
\affil[3]{Toulouse School of Economics, \footnote{21 All\'ee de Brienne, 31000 Toulouse, France} and Capital Fund Management}
\affil[4]{HEC Paris, \footnote{ 1 rue de la Lib\'eration, 78351 Jouy en Josas, France} and CEPR}

\date{\today}
\maketitle
\abstract{This note investigates the causes of the quality anomaly, which is one of the strongest and most scalable anomalies in equity markets. We explore two potential explanations. The ``risk view'', whereby investing in high quality firms is somehow riskier, so that the higher returns of a quality portfolio are a compensation for risk exposure. This view is consistent with the Efficient Market Hypothesis. The other view is the ``behavioral view'', which states that some investors persistently underestimate the true value of high quality firms. We find no evidence in favor of the ``risk view'': The returns from investing in quality firms are abnormally high on a risk-adjusted basis, and are not prone to crashes. We provide novel evidence in favor of the ``behavioral view'': In their forecasts of future prices, and while being overall overoptimistic, analysts systematically underestimate the future return of high quality firms, compared to low quality firms.}

\setcounter{page}{1}


\newpage

\section{Motivation}

\subsection{The Quality Anomaly}

\begin{figure}[h!]
\caption{Cumulative Return of a Quality Anomaly}
\label{perf_ocf}
\begin{center}
\includegraphics[scale=.7]{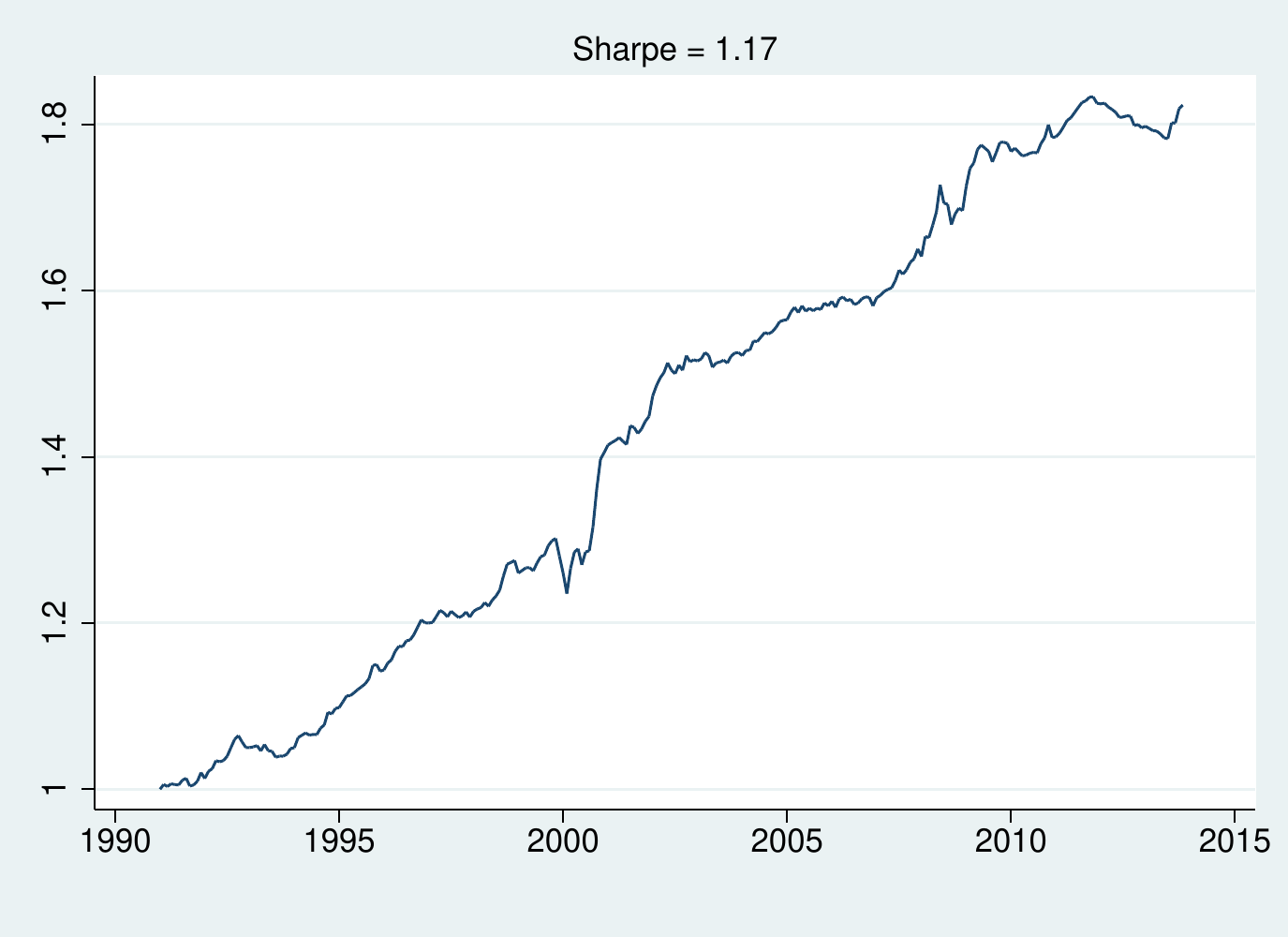}
\end{center}
\begin{footnotesize}
P\&L chart of the quality market-neutral strategy, long US firms with high cash-flows to total assets (COMPUSTAT item OANCF divided by item AT), 
short US firms with high cash-flows to total assets. Details on the data and portfolio construction are given in the Appendix. 
\end{footnotesize}
\end{figure}

The so-called ``quality anomaly" is one of the capital markets' strongest reported anomalies and has a long tradition among investors (\cite{beats}, and more recently \cite{quality} \cite{novy}). 
It chiefly amounts to ranking firms in terms of their ratio of operating cash-flows (OCF) to total assets or alternatively of their returns to total assets (ROA), 
as indicators of the profitability (or quality) of the firms. The portfolio is long high-quality stocks and 
short low-quality stocks, in a market neutral way (see Appendix for more details on the portfolio construction). Figure \ref{perf_ocf} illustrates the past performance of such a strategy for US stocks. 
Quite surprisingly, this performance is quite high: even absent cross-country diversification (this is US data only), one already obtains a Sharpe Ratio of $1.2$ over the period 1990-2012, corresponding to a 
highly significant t-stat $\approx 6$. The same strategy is statistically significant in all geographical zones, 
and the corresponding signal moves sufficiently slowly so that large amounts of capital can be invested without 
suffering from prohibitive transaction costs -- see \citep{LST}. The quality anomaly is therefore economically significant and works surprisingly well compared to other well-documented anomalies. We report the performance of 8 other anomalies in US equity markets in Table \ref{perf and co}. Over the same period (1990-2012), on the same set of stocks, Momentum \citep{santaclara}, Low Vol \citep{ang}, Net Repurchasers \citep{pontiff} and Industry Leaders \citep{hou} 
all have a Sharpe ratio $\approx 0.5$; well below the Sharpe ratio of 1.2 for cash-flows to total assets. In addition, signals such as Industry Leaders or Momentum mean-revert more quickly, leading to larger transaction costs 
and smaller capacities \citep{LST}.

\begin{table}[h!] 
\begin{center}
\caption{Risk-Return Profile of Various Strategies}
\label{perf and co}
\medskip{}
\small
\begin{tabular}{l cccccc }\\
\hline\hline
& (1) & (2) & (3) & (4) & (5) & (6)\\
& Sharpe&  &  &  & Proba  & Signal\\
& Ratio& $\beta $& $\beta^{-}$&$ Skewness $ & $\left(r_t<-2 \sigma \right)$ & Persistence \\
\hline \\
Market - short rate&         .47&           1&           1&        -.13&        .031&           .\\
Low vol     &         .43&       -.015&           0&        -.06&        .032&         .99\\
Book to Market&          .2&        .029&         .11&        .035&        .025&         .98\\
Repurchasers&         .55&         .01&         .04&       -.053&        .019&         .96\\
 Momentum   &         .43&       -.041&         -.1&       -.007&        .025&         .88\\
Industry Leaders&         .48&       -.016&        -.14&        .008&        .029&         .15\\
Accruals    &         .77&        .014&       -.027&        .027&        .018&         .95\\
ROE         &         .55&       -.025&       -.033&        .021&         .01&         .97\\
Cash-Flows  &         1.2&       -.016&       -.055&         .06&        .021&         .97\\
ROA         &         .46&       -.025&       -.054&         .08&         .01&         .99\\
 
\hline \hline
\end{tabular}
\vfill
\end{center}
{\footnotesize This table describes the risk-return profile of 8 well-known quantitative long-short strategies and long the market itself, using monthly returns. See Appendix for 
details on data, signal, and portfolio construction. The first strategy corresponds to the zero cost market portfolio.  The next first 5 strategies correspond to classical strategies, where the signal is: minus the rolling three months daily volatility, the book to market ratio, taking the most recent market value and the last available book value of equity, minus the growth in adjusted shares outstanding, the past cumulative return from month $t-12$ to $t-2$ and the last month's return of the largest stocks in the same 2-digit industry. Then, we report three variations around the quality strategy: return on assets (EBIT/Total assets), return on equity (Net Income/Common equity) and cash-flows (net operating cash-flows/total assets). For each of these hedged strategies, we report the historical annualized Sharpe ratio, the market $\beta$ (which should be small if the beta does not change too much, the $\beta$ on negative market months, a measure of skewness (mean minus median normalized by standard deviation) and the frequency of months with returns lower than $-2 \sigma$. Note that all the quality strategies have a positive skewness. 
The last column reports the average persistence of the monthly signal, which is the $b$ coefficient in the regression $s_{it}=a+b \, s_{it-1}+\epsilon_{it}$.}
\\
\end{table}

\subsection{Risk Premium or Behavioral Anomaly?}

The strength, universality and persistence of the quality anomaly cries for an explanation. After all, the information about operating cash-flows is public and can hardly be seen as complex or difficult to process. A possibility -- consistent with the Efficient Markets Hypothesis (EMH) -- is that these excess returns would be somehow related to a risk premium. High quality stocks could be inherently exposed to a risk factor that investors care about. One possible story could be the following: firms can choose between safe and moderately profitable projects, and risky and more profitable ones. High average cash-flow to assets might indicate that the corresponding firms are 
themselves, on average, operating on very profitable but riskier segments of the economy and therefore riskier to own. Alternative stories have been proposed, based on the fact that investments cannot be costlessly reversed (see e.g. \cite{zangzang}, \cite{kisser}). However, while well-known risk premia strategies are indeed rewarding investors for carrying a significant negative skewness risk (see e.g. \cite{harveysiddique}, \cite{lemperiere}), quality strategies are in fact found to have a {\it positive} skewness and a very small propensity 
to crash -- see  Table \ref{perf and co}. This makes the risk premium interpretation very unlikely.

A more plausible interpretation, at least to our eyes, is based on psychological biases and is inconsistent with the EMH. The behavioral view posits that investors systematically underweight the information contained in quality-like signals (cash-flows, ROA, etc.). A possible story is misplaced focus: analysts and investors are too focused on other indicators such as Earning per Share (EPS), Momentum, volatility, etc. and fail to use all the information available in financial statements. Since accounting conventions that lead to the calculation of EPS paint a less reliable picture than the cash-flow statement \citep{sloan}, cash-flows or ROA contain information that is relevant to value companies but might be insufficiently used by analysts and investors. A possibly complementary story, fully explored in our companion paper \cite{BKLT}, is conservatism bias: the agents' beliefs are ``sticky'' and this leads to under-reaction to good or bad news about the firms. 

In this note, we run statistical tests on an analyst recommendation data base (IBES) and find strong support for the behavioral view. We focus here on analysts' price targets, and regress their anticipation of the future return 
of a firm on its quality (measured by cash-flows over assets). While the intercept is positive, indicative of a systematic over-optimism already reported in the literature, the slope of the regression is found to be {\it negative} at the 5\% significance level. This shows that analysts, at best, neglect the information contained in cash-flow statements -- or even weight it with the wrong sign. In \cite{BKLT}, we study expected earnings by analysts (rather than expected prices) and relate the ``stickiness" of analysts' beliefs to asset mispricing. We believe that these two widespread behavioral biases (misplaced focus and ``stickiness") are at the heart of the quality anomaly.

\section{The Behavioral Origin of the Quality Anomaly}

\subsection{Expectation biases}

The psychology literature documents many instances where humans tend to form biased beliefs. This does not necessarily mean that their beliefs are random, but that the (average) direction of their mistakes can be predicted. For instance, opinions often tend to be too sticky, as people update them in a slower fashion than that implied by Bayes' rule. This ``conservatism bias'' is documented using forecasts by experts in \cite{coibion} and is often refered to as ``under-reaction" in the finance literature. The psychology literature documents that mistakes can be self-reinforcing and persistent as agents tend to discard new information that is not congruent with their priors, a phenomenon called ``confirmatory bias'' \citep{rabin}, which also leads to conservatism.  People also tend to overestimate the quality of their private information (``overconfidence''), which makes them insufficiently take into account relevant pieces of public information.  In other contexts, agents over-react as they overweight information that is highly salient (\cite{griffin}, \cite{debondt}). In the context of finance, investors might focus too much on salient  information, say EPS growth or momentum, and neglect other informative signals about company performance (cash-flows, accruals, etc.). Due to these biased beliefs, stock prices might take time to fully reflect existing information on companies.

One can test whether a group of agents has biased beliefs provided one observes a large set of beliefs of these agents. Assume that we can observe the forecast at time $t$ of future stock returns 
(realized by holding the stock between time $t$ and $t+1$) by a given investor regarding company $i \in I$: $F R_{i, t+1}$. The letter $F$ stands for forecast, as opposed to $E R_{i, t+1}$ which is the ``rational'' unbiased expectation  of prices of company $i$. The expectation bias of this investor is the gap between her subjective expectations 
and the rational expectations of future yearly returns and is given by:
$$
\text{bias}=F R_{i,t+1} - E R_{i,t+1} = \left[ F R_{i,t+1} -  R_{i,t+1}\right] - \left[ E R_{i,t+1} - R_{i,t+1} \right],  
$$
which is decomposed into two pieces.  The first term in brackets measures the difference between forecasts (subjective expectations) and realizations. The second term in squared brackets is the ``rational'' expectation error. 
It comes from the fact that even the savviest investors cannot forecast the future: there is uncertainty that cannot be eliminated. The difference between the two terms is the contribution of the systematic bias of investors 
to their mistake. The rational expectation $E R_{i,t+1}$ is not directly observable but we can use the fact that on average over a large number of observations, it should be identical to the average realization of $R_{i,t+1}$ 
(i.e. the average of the second term in brackets is zero).

To see this more formally: when we average the equation above on a large set of companies $i \in I$ and time-periods $t \in T$ conditional on firm characteristics $X$ (for example $X=\text{OCF}$), we can take advantage of the fact that: 
$$ 
E_{\substack{i \in I,t \in T}} \left[E R_{i,t+1} -  R_{i,t+1}\right | X] =0,
$$ 
and get an expression for the bias of subjective expectation vis-a-vis rational expectations, conditional on any firm characteristics $X$:
$$
\text{bias}(X) = E_{\substack{i \in I,t \in T}}  \left[  F R_{i,t+1} -  R_{t+1}  \right | X]
$$
To check whether investors or analysts tend to be over-optimistic or pessimistic regarding companies that have a given set of characteristics, we can check whether her errors are bigger or smaller on average for companies that have those characteristics, i.e. test whether bias$(X)$ is not only statistically different from $0$, but in fact depends
on the specific value of $X$.  By taking the average over a sufficiently large number of companies, we remove the rational mistake term, and are able to directly measure the bias 
(by definition, the rational expectation $E R_{t+1}$ is not biased, so that it is equal to the average realized $R_{t+1}$).

\subsection{Analysts Analyzed}

Whether agents have beliefs that are biased or not is often hard to prove as beliefs are rarely directly observable. In the context of finance, we luckily do have information on the beliefs of a relevant set of market participants: company analysts. Analysts regularly issue and update forecasts about future earnings and future stock-prices of companies they follow. If they were ``rational", their forecast errors, i.e. the difference between realized and forecast values, should not be predictable based on information available at the time of the forecast. But if analysts tend to underweight some information (e.g. cash-flows to assets), we should see that the direction and size of their mistake directly depends on such neglected information. A few papers have already used analyst forecasts in order to explore biases in expectations, but very few are actually looking at the cross-section of expectation \emph{mistakes} to reject the rational expectation hypothesis. \cite{laporta} and \cite{michaely} focus on expectations and implement indirect tests. \cite{laporta} finds that firms for which analysts are bullish (have high growth expectations) tend to underperform in the medium run, which indicates that they are overoptimistic. \cite{michaely} finds that analyst price forecasts tend to be lower among value firms, which suggests that they are pessimistic, but they do not compare expected and realized prices. \cite{gennaioli} show that optimistic managers tend to invest more, controlling for classical determinants of investment. Closer to the present note, because they actually use expectation mistakes, \cite{Entrepreneurs} find that entrepreneurs that are persistently optimistic compared to realized outcomes take more short-term debt.

To measure expectations, we use IBES data in the period 2003-2012 and focus on analyst forecasts of stock prices (for a related analysis using earning forecasts, see our companion paper \cite{BKLT}). 
Specifically, we retrieve from IBES the ``target price'' that analysts report for the firms they follow. The target price is the price that analysts expect that the stock will reach within one year. At each month end, we compute the average target price across all forecasts that are active. The average target price $F P_{t+12}$ allows us to compute forecasted returns as: $F R_{t+12}=F P_{t+12}/P_{t}-1.$ 
Our time unit is monthly and forecasts all have a 12 months horizon. Following the above discussion, we compute at the end of each month $t$ expectation mistakes using the formula:
\begin{equation}
\label{jules romain}
\text{mistake} =F R_{t+12} - R_{t+12}
\end{equation}
The distribution of forecasting mistakes is found to be well behaved, without many outliers. The average expectation mistake is +.08 (median $=$ .06) which is proof that analysts are on average slightly optimistic. A large literature has documented such a bias for EPS forecasts (see e.g. \cite{guedj}). It can be attributed, for example, to conflicts of interest: Most analysts have an incentive to recommend that the stocks be bought rather than sold. 

\subsection{Regressing the Bias on Quality}

However, our approach does not consist in looking at the average bias, which is, for the reasons listed above, to be expected positive and not very informative. Instead, we compare in the cross-section, the bias between high and low quality firms. If the bias is larger (more positive) among low quality firms, we will interpret it as evidence that investors are relatively optimistic (pessimistic) about low quality (high quality) firms. Our statement is a relative one as we believe it is easier to interpret the difference in biases across categories of stocks, rather than their average value. 

We implement our statistical test by running the following linear regression, for stock $i$ in year $t$:
$$ 
\text{mistake}_{i,t} = \beta \text{Quality}_{i,t}+ \text{controls}_{i,t} +\epsilon_{i,t} 
$$
where $\text{Quality}_{i,t}$ stands for one quality quantifier, such as OCF or ROA, and $\text{controls}_{i,t}$ are various firm characteristics that may affect the expectation mistake. We allow for flexible correlation structure for all shocks $\epsilon$ across observations of each stock (mistakes can be autocorrelated, it will not affect our estimates). We also include one dummy per date (time fixed effect) to ensure that ($\beta$) is estimated through the difference in expectation errors for high vs. low quality firms in a given month.

We report the results of this regression in Table \ref{jpb} using last available cash-flows to total assets as our proxy for firm quality (we also used EBIT to assets and Earnings to Equity and find very similar results). We are careful to avoid ``look-ahead bias'', by making sure that the firm-level quality variable used at time t is only based on informations that are public as of time t. We rank the firms at each time t according to this measure of quality, and we normalize the rank so that it goes from -.5 to .5 (we do this to deal with outliers, but our findings do not depend on this particular normalization). We apply a similar normalization to all control variables. In each column, we report the result of one regression: in front of each variable, we show the coefficient of this variable in the regression of the column. Columns (1), (3) and (5) do not include controls, they measure the raw correlation between expectation mistake and quality. We find that analysts on average make significantly lower mistakes on high quality firms, i.e. they are relatively more optimistic about low quality firms. The coefficients of the regression can be directly interpreted as our Operating Cash Flows variable is normalized to be uniform on $[-.5,.5]$: a company that is at the top of quality distribution has an unexpected positive over-performance of 6\%  yearly vis-a-vis a company that is at the bottom of the distribution. This finding carries out when we control for other potential drivers of expectation mistakes: we include Book to Market and past realized volatility. Interestingly, it seems that analysts tend to be too optimistic about growth stocks (those that have low book-to-market) and high vol stocks. This suggests that the low vol and value anomalies might also have behavioral origins rooted in expectations mistakes, like quality. (For more on the low vol anomaly, see e.g. \cite{BakerWurgler}, \cite{BakerW}, \cite{lowvol} and refs. therein). More importantly, the fact that the coefficient on quality is unaffected in magnitude and statistical significance by the controls suggests that the link between forecast mistakes and quality is not due to the volatility or growth characteristics of the stocks.

\begin{table}[htbp] 
\begin{center}
\caption{Realized vs. Expected Yearly Returns: Analysts Underweight Quality}
\label{jpb}
\medskip{}
\small
\begin{tabular}{l cc | cc | cc }\\
\hline\hline
&\multicolumn{2}{c}{Mistake}& \multicolumn{2}{c}{Forecast} & \multicolumn{2}{c}{Realized} \\ \\
& (1) & (2) & (3) & (4) & (5) & (6) \\
\hline 
&&&&&& \\
Op. Cash Flows      &       -.063***&       -.069***&       -.012** &       -.005   &         .05***&        .064***\\
                    &      (-6.2)   &      (-6.4)   &      (-2.4)   &      (-1.1)   &       (6.5)   &       (7.1)   \\
Rolling volatility  &               &         .14***&               &         .13***&               &      -.0075   \\
                    &               &        (14)   &               &        (32)   &               &        (-1)   \\
Book to Market      &               &       -.044***&               &       -.011** &               &        .033***\\
                    &               &      (-3.8)   &               &      (-2.5)   &               &       (3.6)   \\
r2                  &         .27   &         .28   &         .24   &         .29   &         .26   &         .27   \\
N                   &      136967   &      133917   &      136967   &      133917   &      148975   &      145486   \\
 
&&&&&& \\
Month FE & YES & YES & YES & YES & YES & YES \\
Cluster & Firm & Firm & Firm & Firm & Firm & Firm \\
\hline \hline
\end{tabular}
\vfill
\end{center}
{\footnotesize Note: The data cover the 2003-2012 period. Each column corresponds to one regression. In the first two columns the left-hand side is the ``mistake'' (i.e. expected returns by analysts (consensus) over the next 12 months minus the corresponding realized returns), as a function of firm characteristics observable at the time of forecast. Colums 3-4 run the same regressions for forecasted returns and columns 5-6 for realized returns. The 
forecast results are specially meaningful for our behavioral interpretation.
$^{***}$ means ``statistically significant at 1\%'', $^{**}$ means ``statistically significant at 5\%''. Standard errors are clustered at the firm level.
}
\\
\end{table}

\subsection{Separating Performance from Forecasts}

Although interesting, the above analysis in fact mixes two different effects. One is nearly tautologically associated to the very existence of the quality anomaly: regressing the future realized return on Quality must lead to 
a positive regression coefficient if the strategy is to be profitable. For our purpose, it is however also meaningful to regress the analysts' forecast itself, as a way to check directly how much analysts underweight the signal in their expectations. We thus re-run separate regressions for realized and forecasted returns as a function of Quality. 

The regression coefficients are given in Table \ref{jpb}, columns 3-6, and constitute the most striking result of the present study: analysts are clearly under-weighting operating cash-flows; in fact they even seem to be putting a slightly negative weight on that variable, even though it is a strong positive predictor of future returns. 

These last results strongly suggest that the quality anomaly is likely due to a significant underweighting of quality in price forecasts. For this 
to transpire in market prices, one needs to further assume that analysts have some influence on other market participants, or simply if their opinions are  representative of those of investors. We regard this hypothesis to be quite plausible.

\section{Conclusion}

In this note, we explored the potential explanations for the quality anomaly, which is one of the strongest in equity markets. Since the returns of a strategy long the most profitable companies and short the less profitable ones has a {\it positive} skewness, it is hard to account for this anomaly using a risk premium argument, which are characterized by a negative skewness (i.e. a propensity to crash). 

By contrast, we report strong evidence that financial analysts pay insufficient attention -- or even negative attention -- to firm-level profitability indicators such as Operating Cash-Flows, leading to forecast errors that are negatively correlated with quality indicators. This suggests that the quality anomaly arises from non-optimal weighting of profitability information by analysts and investors in their expectations. This behavioral bias might be due to relative excess attention to more salient accounting information, or, as documented in our companion study \cite{BKLT}, to under-reaction to positive (or negative) news about the firm. This under-reaction is present in many other contexts, for example in the behaviour of macroeconomic experts, see \cite{coibion}, with time-scales that are surprisingly similar to the ones observed for financial analysts (see \cite{BKLT}). 
Since the conservatism bias could also be -- at least partly -- responsible for the appearance of trends in financial markets (see \cite{BKLT}), we claim that this bias is probably the most persistent and economically significant of all documented behavioral biases. How long investors will (on aggregate) remain inattentive and sticky in an era more and more dominated by computerized data analysis remains to be seen.

\section*{Appendix: Data \& Portfolio construction}

The sample used is constituted of the largest 1,500 stocks on US markets over the 1990-2012 period. The data source is CRSP and COMPUSTAT.
Each market-neutral strategy is constructed as follows: the ``signal'' (e.g. OCF) is ranked from -.5 to +.5, and this rank is the portfolio weight. 
The portfolio is then hedged by shorting $\beta_t$ \$ of market per \$ of long exposure, where $\beta_t$ is the rolling beta of the portfolio over the past 24 months. The strategy's returns are measured per unit of gross value of the hedged portfolio (i.e. profits divided by the long market value plus short market value of the hedged portfolio). It is crucial in back-testing exercises that information used to form portfolios be available at time of portfolio formation. To be conservative, we sort firms on ``old'' values of the signal. More precisely, we take the last available annual accounts if the last fiscal year has ended at least 6 months ago, or the penultimate annual accounts if the last fiscal year has ended less than 6 months ago (Annual reports typically take at most 3 months until publication). Due to the use of rolling betas, this dynamic hedging procedure has no ``look-ahead bias'': all information used to hedge the portfolio in a given month is available at the beginning of that month.


\newpage
\bibliography{WhyQualitybib}
\bibliographystyle{aer}

\end{document}